\documentclass[aps,twocolumn, superscriptaddress, nofootinbib]{revtex4-1}

\usepackage{adjustbox}
\usepackage[usenames]{color}
\usepackage{xcolor}
\usepackage{upgreek}
\usepackage{amsmath}
\usepackage{amssymb} 
\usepackage{graphicx}
\usepackage[utf8]{inputenc}
\usepackage{physics}
\usepackage{empheq}
\usepackage{braket}
\usepackage{siunitx}
\usepackage{float}
\usepackage{tcolorbox}
\usepackage{enumitem}
\usepackage{placeins}
\usepackage{xcolor}
\usepackage[T1]{fontenc}
\DeclareMathSizes{10}{10}{10}{10}

\newcommand{\ah}{\hat{a}}
\newcommand{\bh}{\hat{b}}

\begin{document}
\title{Racetrack computing with a topological boundary ratchet} 
\author{Parisa Omidvar}
\affiliation{AMOLF, Science Park 104, 1098 XG Amsterdam, the Netherlands}
\author{Markus Bestler}
\affiliation{Department of Physics, University of Konstanz, D-78457 Konstanz, Germany}
\author{Sima Zahedi Fard}
\affiliation{AMOLF, Science Park 104, 1098 XG Amsterdam, the Netherlands}
\author{Oded Zilberberg}
\affiliation{Department of Physics, University of Konstanz, D-78457 Konstanz, Germany}
\author{Marc Serra-Garcia}
\affiliation{AMOLF, Science Park 104, 1098 XG Amsterdam, the Netherlands}

\date{\today}

\maketitle
\textbf{Multistable order parameters provide a natural means of encoding non-volatile information in spatial domains, a concept that forms the foundation of magnetic memory devices. However, this stability inherently conflicts with the need to move information around the device for processing and readout. While in magnetic systems, domains can be transported using currents or external fields, mechanisms to robustly shuttle information-bearing domains across neutral systems are scarce. Here, we experimentally realize a topological boundary ratchet in an elastic metamaterial, where digital information is encoded in buckling domains and transported in a quantized manner via cyclic loading. The transport is topological in origin: neighboring domains act as  different topological pumps for their  Bogoliubov excitations, so their interface hosts topological boundary modes. Cyclic loading renders these modes unstable through inter-domain pressure, which in turn drives the motion of the domain wall. We demonstrate that the direction of information propagation can be controlled through adjustable mechanical constraints on the buckling beams, and numerically investigate buckling-based domain-wall logic circuits in an elastic metamaterial network. The  underlying tight-binding structure with low-order nonlinearities makes this approach a general pathway toward racetrack memories in neutral systems. }
\begin{figure}[htbp]
    \centering
    \includegraphics[width=\columnwidth]{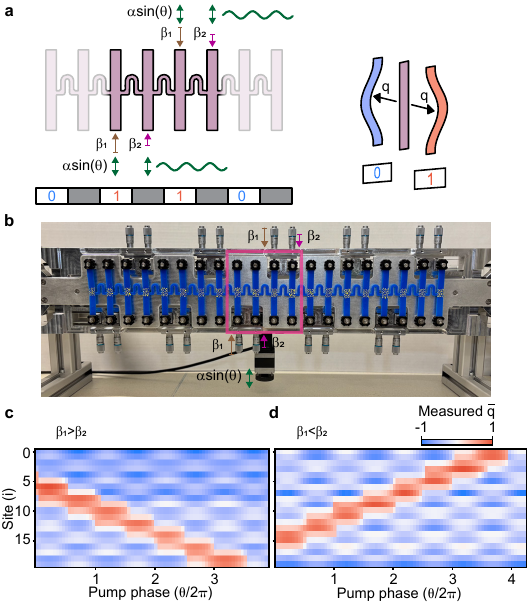}
    \caption{\textbf{ Observation of quantized information transport in an elastic racetrack memory}. \textbf{a}. Sketch of the metamaterial consisting of beams coupled with arced connections. The beams are subject to a constant (DC) and a time-varying (AC)  compression (pink and green arrows, respectively). Every four sites, the compression pattern repeats, forming a unit cell. The beams under higher compression are bistable most of the time and have two stable buckling shapes (red and blue). These two configurations can represent binary digits 0 and 1, allowing the beams to store information like bits in a memory array. \textbf{b}. Polymer sheet assembled on the setup, the pink rectangle marks the unit cell of the structure. \textbf{c,d}. Measured normalized displacements at the centers of the beams ($\overline{q}$) as the AC compression is varied, for two different DC compression $\beta_j$. The pumping starts from a configuration containing one soliton. After a full cycle of the external compression pattern, the soliton is shifted by one unit cell.} 
    \label{fig:SystemAndPropagation}
\end{figure}

The separation of memory and computation, known as the von Neumann bottleneck~\cite{backus1978can}, represents a major inefficiency in modern computer architectures and has been estimated to account for up to 90\% of the energy cost of contemporary AI training~\cite{energyMemory}. Efforts to overcome this bottleneck by integrating memory and computation have motivated the development of racetrack memories~\cite{hayashi2008current, parkin2008magnetic, heydermanDomainWallLogic, pham2024fast, gu2022three, szameit2024discrete}. In these devices, information is stored in magnetic domains or skyrmions and can be transported and manipulated using injected currents or external fields. Moving to neutral, non-magnetic systems, such coupling to external fields is missing, and topological pumping~\cite{ rice1982elementary,thouless1983quantization, kraus2012topological,lohse2016thouless, citro2023thouless} arose as a mechanism for controllable information transport. However, such pumps adiabatically move linear excitations, which are short-lived~\cite{xia2021experimental, jurgensen2021quantized, jurgensen2023quantized}. Hence, mechanisms to transport persistently stored information in a controllable way are still missing.



Our experiment involves a nonlinear elastic metamaterial~\cite{kwakernaak2023counting, fang2025large, bordiga2024automated, ducarme2025exotic, florijn2014programmable} that persistently stores information, and robustly transports it under an external cyclic load. This process is driven by sequential instabilities occurring in localized modes at the boundaries between spatial domains.  
Our racetrack memory is a chain of interconnected beams (Figs. ~\ref{fig:SystemAndPropagation}\textbf{a},\textbf{b}), which were cut from a thin polymer sheet. We create a sublattice potential on the chain by applying a strong compressive displacement $\beta_1$ at one end of every alternate beam. These beams exhibit two stable buckling states, labeled as $0$ and $1$, which can store information. Beams experiencing lower compression $\beta_2$ do not store information, and serve instead as variable couplings. We refer to the beams under high compression as main beams and those under low compression as coupling beams. The racetrack can support different spatial domains, where a domain corresponds to a region in the chain with all the main beams uniformly assuming the values of $0$ or $1$. Between these regions, domain walls appear. By soliton, we dub a domain of $1$ surrounded by two domains of $0$.

To introduce directed soliton motion, we apply an additional compressive displacement in the beams, of the form $\pm\alpha\sin \left[\theta(t)\right]$ (Fig.~\ref{fig:SystemAndPropagation}\textbf{a}). This force alternates sign every two sites, being positive in one pair of coupling-main beams, and negative in the next. This time-dependent compression spatially repeats every four sites, defining a four-site unit cell. When the lattice is modulated, by increasing $\theta$ over time, domain walls and therefore solitons are transported through the system (Fig.~\ref{fig:SystemAndPropagation}\textbf{c}), demonstrating the elastic racetrack memory functionality. If the static compressive displacements $\beta_1$ and $\beta_2$ are exchanged, the direction of propagation reverses (Fig.~\ref{fig:SystemAndPropagation}\textbf{d}), providing a mechanism to read out the information by changing the boundary conditions. In our experiment, the time-dependent compression is applied by clamping the alternating ends of the beams to a moving plate, that is driven by a single motor (Fig.~\ref{fig:SystemAndPropagation}\textbf{b}).



\begin{figure}[htbp]
    \centering
    \includegraphics[width=\columnwidth]{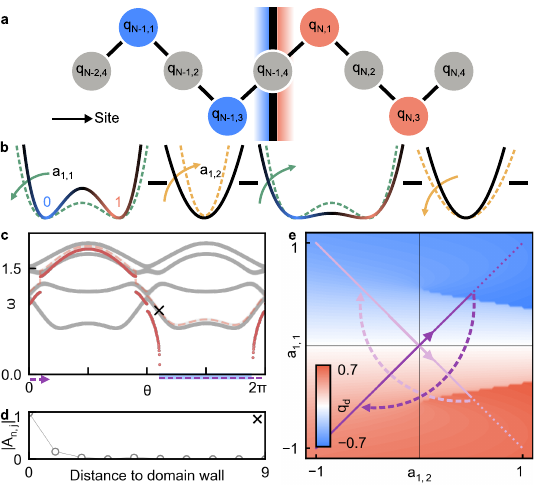}
    \caption{\textbf{Theory of Soliton Movement.}
    \textbf{a} Sketch of the tight-binding model corresponding to the experiment in Figs.~\ref{fig:SystemAndPropagation}\textbf{a}, described by Eq.~\eqref{eq:GrossPitaevskii}, including a domain wall (black vertical line in the middle). Grey circles stand for monostable sites while red (blue) circles depict bistable sites initialized in the 1 (0) state.
    \textbf{b}
    Onsite potentials (black lines) of the 4 sites of the unit cell as $\theta$ is varied. The effect of larger $a_{1,1}$ ($a_{1,2}$) on the main (coupling) beams is marked as green (yellow) dashed lines. Horizontal black lines display coupling between the potentials.
    \textbf{c} Band structure (grey) of Bogoliubov excitations on top of the red domain from \textbf{a} during one pump cycle of $\theta$, with $a_{1,1}=a_{1,2}=-\sin(\theta)$ for a racetrack of 80 sites with $\lambda=1$, $c=0.25$, $\omega_0^2=2$, $a_{0,1}=-3$, and $a_{0,2}=-0.5$. Boundary mode frequencies when coupled to vacuum (dashed red) and to the second domain (red dots). Regions where the frequencies coupled to the second domain turn imaginary (unstable) are marked in blue. Encountering instabilities shifts $\theta$ by $\pi$ (purple dashed arrow).
    \textbf{d} Mode shape $|A_{n,j}|$ (see Appendix) for a localized fluctuation mode of the unperturbed red domain from \textbf{a} for 10 sites and $\theta\approx1.1\pi$ marked with a black cross in \textbf{c}. Site 0 corresponds to site $q_{N,1}$.
    \textbf{e} Instability phase diagram for same parameters as in \textbf{c} for a racetrack of 16 sites. Red (blue) regions indicate where the blue (red) domain becomes unstable. The dark purple line shows the pump trajectory in our system, where $\theta$ advances by $\pi$ when reaching an instability due to modified boundary conditions (dashed arrow). Hence, the pump cycle never reaches a part of the pump cycle (dotted line). Exchanging $\beta_1$ and $\beta_2$ effectively leads to a new pump cycle on top of the instability diagram (light purple).}
    \label{fig:Fig2TheoryOfSoliton}
\end{figure}

Our soliton displacement stems from the motion of the two domain walls defining it. Hence, we turn to investigate the mechanism driving the domain wall motion. Our elastic racetrack memory (Fig.~\ref{fig:SystemAndPropagation}\textbf{a}) is a nonlinear perturbative metamaterial~\cite{matlack2018designing, EmbodyingSima2025} governed by the tight-binding potential (Fig.~\ref{fig:Fig2TheoryOfSoliton}\textbf{a}) 

%
\begin{equation}
\begin{split}
V &=\sum_{n,j} [\frac{\lambda}{4}q^4_{n,j}+\frac{1}{2}(\omega_0^2+a_{0,j}+a_{1,j}(\theta))q_{n,j}^2]  \\
&  + \sum_{l}\frac{c}{2}(q_l-q_{l+1})^2,
\end{split}
\label{eq:GrossPitaevskii}
\end{equation}
with center displacement $q_{n,j}$ of the $j$-th beam ($j\in\{1,2,3,4\}$) in the $n$-th unit cell, quartic nonlinearity $\lambda$, and harmonic potential with natural frequency $\omega_0$. The system is spatially modulated by the constant compressive displacements $\beta$, which introduce tight-binding potential terms $a_{0,j}=\pm\gamma\beta$ due to geometric nonlinearity, quantified by $\gamma$~\cite{serra2016mechanical}. Similarly, the pumping phase ($\theta$)-dependent compressive displacements introduce tight-binding potential terms of the form $a_{1,j}(\theta)=\pm\gamma\alpha\sin (\theta)$. Figure~\ref{fig:Fig2TheoryOfSoliton}\textbf{b} shows the resulting modulation of the local potential as $\theta$ is varied. Nearest neighbors are attractively coupled with constant $c>0$. For compact notation, we let $l$ index all pairs of $i$ and $j$, with $l+1$ corresponding to the next site in the racetrack. The static compressive displacements are chosen such that $a_{0,1}=a_{0,3}$ and $a_{0,2}=a_{0,4}$. The quadratic term of each site, $\frac{1}{2}(a_0 + a_{1}(\theta) + \omega_0^2 + 2c)q_{i}^2$, is negative (positive) for the first and third (second and fourth) site of the unit cell. As a result, main (bistable) and coupling (monostable) sites alternate along the racetrack. In the experiment, we vary the compressive displacements with the pumping phase such that $a_{1,1}=-a_{1,3}$ and $a_{1,2}=-a_{1,4}$ (cf.~Figs.~\ref{fig:SystemAndPropagation}\textbf{a} and \textbf{b} with Fig.~\ref{fig:Fig2TheoryOfSoliton}\textbf{b}). In the following theory analysis, we first assume that $a_{1,1}$ and $a_{1,2}$ can be independently tuned.

The metamaterial model~\eqref{eq:GrossPitaevskii} can be viewed as a nonlinear extension of a topological pump, owing to the four-site periodic modulation of the local harmonic potential~\cite{thouless1983quantization, kraus2012topological,lohse2016thouless, citro2023thouless}. On a single domain, the linearized Bogoliubov excitation modes can acquire a nontrivial Chern number when $a_{1,1}(\theta)$ and $a_{1,2}(\theta)$ are varied independently (see Appendix). One might therefore expect the quantized soliton motion to originate from the topology of the Bogoliubov modes, cf.~Refs.~\cite{Jurgensen2022ChernNumber, mostaan2022quantized}. However, when $a_{1,1}(\theta)$ and $a_{1,2}(\theta)$ are tuned simultaneously as in the experiment, the Bogoliubov bulk bands close during the pump cycle and the Chern number becomes ill-defined, see Fig.~\ref{fig:Fig2TheoryOfSoliton}\textbf{c}. Nonetheless, at points along the pump cycle, where the excitation gaps remain open, boundary modes still manifest in the gaps (Fig.~\ref{fig:Fig2TheoryOfSoliton}\textbf{d}) due to nonvanishing bulk polarization that develops as $\theta$ is varied.

To trace the origin of soliton motion, we numerically simulate a system with a single domain wall (Fig.~\ref{fig:Fig2TheoryOfSoliton}\textbf{a}). For each point in the $a_{1,1}$–$a_{1,2}$ plane, the system is initialized in a single-domain-wall state and allowed to relax to equilibrium. The wall displacement is quantified by $q_\mathrm{d}=(q_{N-1,3}+q_{N,1})/2$ (Fig.~\ref{fig:Fig2TheoryOfSoliton}\textbf{e}). Values near zero indicate stability, while values approaching $\pm0.7$ mark domain wall motion: the wall relaxes to a shifted single-wall state, with the sign of $q_\mathrm{d}$ determining the direction. This identifies a so-called \emph{topological ratchet} mechanism~\cite{theorypaper}, where wall motion is triggered by instabilities rather than by Chern number transport. Consequently, although the experimental scan of $\theta$ traces a straight line in the $a_{1,1}$–$a_{1,2}$ plane (with bulk-band closings at the origin), it encounters only a single instability, enforcing unidirectional wall motion. This yields quantized transport controlled by a single parameter. Furthermore,  reversing the boundary conditions ($\beta_1$ and $\beta_2$) as in Figs.~\ref{fig:SystemAndPropagation}\textbf{c,d} interchanges main and coupling sites. In tight-binding language, this maps $a_{0,1}\rightarrow a_{0,2}$ and $a_{0,2}\rightarrow a_{0,1}$ and mirrors the unit cell. We therefore reverse the propagation direction of the domain wall.

Although distinct from a conventional topological pump, the ratchet mechanism remains topologically rooted. Analyzing excitations on a racetrack with a domain wall reveals boundary modes localized at the inter-domain boundary as $\theta$ varies. Unlike a vacuum edge, the opposing domain exerts an additional “pressure” \cite{Kellendonk2004BoundaryPressure}, which can destabilize the boundary mode (see Appendix and Ref.~\cite{theorypaper}). As shown in Fig.~\ref{fig:Fig2TheoryOfSoliton}\textbf{c}, this pressure softens the boundary-mode frequency until it becomes imaginary, at which point the wall shifts by two sites (one main beam, one coupling beam). Each displacement of the domain wall modifies its boundary condition, equivalent to advancing the pumping phase $\theta$ by $\pi$ (Figs.~\ref{fig:Fig2TheoryOfSoliton}\textbf{c,e})~\cite{kraus2012topological,theorypaper}. We thus provide an additional way to probe whether the ratchet-like instability is the mechanism driving the observed directed domain-wall motion.

\begin{figure}[tbp]
    \centering
    \includegraphics[width=\columnwidth]{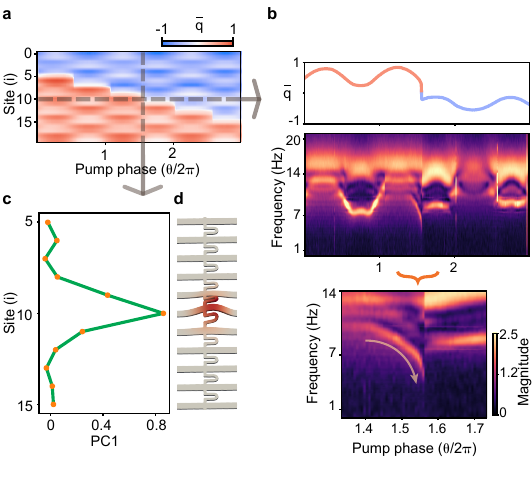}
    \caption{\textbf{Experimental imaging of the Bogoliubov spectrum during domain wall propagation.} 
    \textbf{a} Measured normalized displacements of the beams with a varying compression displacement. The initial configuration of the sample consists of a single domain wall that propagates throughout the experiment. 
    \textbf{b} Buckling displacement (top) and the frequency response (bottom) of the 10th beam in different stages of the experiment. The inset zooms in on the region when the beam transitions (or "hops") between its two stable states. At this moment, the frequency of the mode drops close to zero.
    \textbf{c} Mode shape imaging of the structure at the moment of the 10th beam's transition, obtained using Principal Component Analysis (PCA). The mode is strongly localized at the transitioning beam, confirming that the destabilization corresponds to a localized mode, which allows the domain wall to move forward.
    \textbf{d} First mode shape of the system, simulated using the Finite Element Method (FEM) at a selected moment corresponding to the 10th beam's transition. The simulation shows the same localization as observed in \textbf{c}.}
    \label{fig:experimentalImaging}
\end{figure} 


\begin{figure*}[t]

    \includegraphics[width=\textwidth]{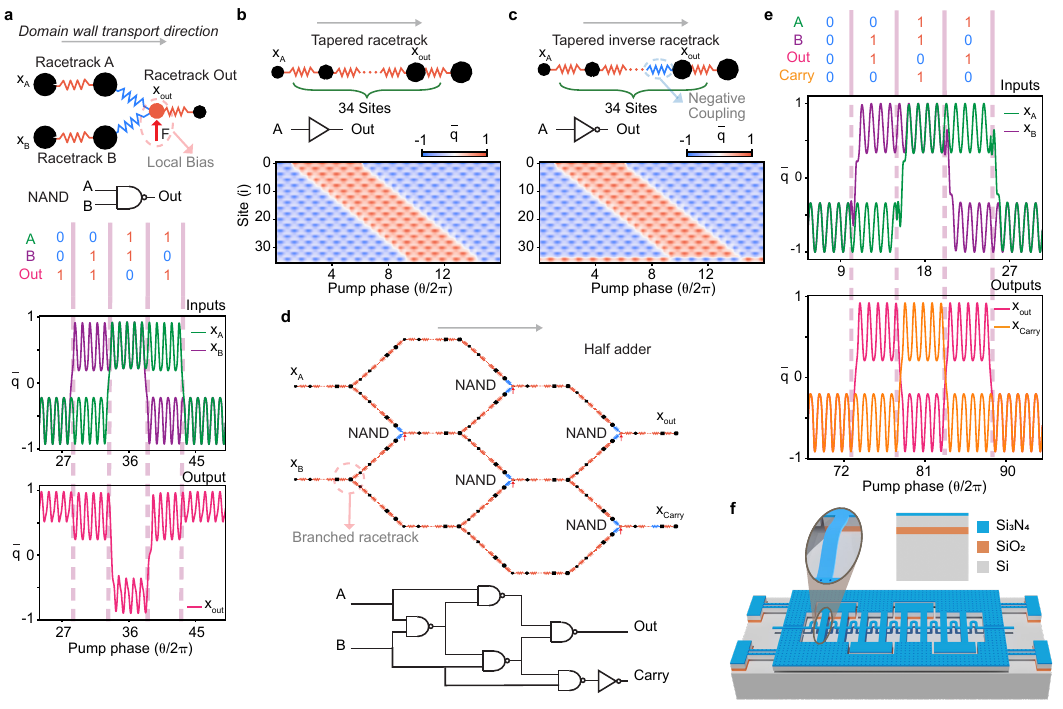}
    \caption{\textbf{Numerical investigation of domain wall logic.} \textbf{a} Simulation of a NAND logic gate using the unit cell in Fig.~\ref{fig:Fig2TheoryOfSoliton}\textbf{b}. The gate has two input racetrack (A, B) with a softer output racetrack. The local positive force breaks the symmetry of output's potential (the red arrow), biasing it toward state $1$. The blue springs represent effective negative coupling. The two graphs show the computed normalized displacement of output and inputs, accurately reflecting NAND logic. The positive displacement values correspond to logical $1$, and the negative values correspond to logical $0$.
    \textbf{b} Top: Tapered racetrack design enabling signal transfer from a softer input site to a stiffer output site, which functions as buffer and amplifier. Bottom: Signal propagation through the racetrack.
    \textbf{c} Top: Tapered inverse racetrack that inverts the transmitted signal from the input to the output. This is achieved by using an effective negative coupling between the input and output sites, functioning as logical NOT gate. Bottom: Inverted signal propagation in the racetrack.
    \textbf{d} Top: Honeycomb-shaped design of a half adder constructed from cascading the elastic logic elements, including buffer, NOT and NAND gates. To form the honeycomb layout, branching is introduced at selected points along the racetracks. Bottom: The corresponding symbolic logic diagram.
    \textbf{e} Truth table of a half adder accompanied by the graphs illustrating computed normalized displacements of the input and the output sites in the honeycomb-shaped design. The half adder takes two one-digit binary inputs (A and B) and adds them into a two-digit binary output representing the sum (out) and carry.
    \textbf{f} 3D sketch of the device layout on a silicon wafer. This conceptual sketch highlights that if the device is miniaturized at microscale, the layout remains planar and fabrication-friendly.
    }
    \label{fig:logic}
\end{figure*}

To test this hypothesis, we designed a new experiment to probe and observe the destabilization of the localized boundary mode. We initialize the metamaterial in a configuration containing two domains separated by a domain wall (Fig.~\ref{fig:experimentalImaging}\textbf{a}). Then, we measure the fluctuation spectrum at the central beam ($i=10$), by introducing small broadband excitations into the system and Fourier-transforming the resulting beam vibration (See Supplementary Information for experimental details). During the measurement, we continuously increase $\theta$, causing the domain wall to propagate. The results are shown in Fig.~\ref{fig:experimentalImaging}\textbf{b}. When the domain wall is far away from the central beam, the spectrum displays the Bogoliubov bulk bands on top of the domain where the central beam resides; their frequencies are modulated by the variable compression of the beam. However, when the domain wall approaches the central beam, we observe an additional peak in the spectrum, whose frequency goes to zero right before the domain wall advances. A modal frequency going to zero is indicative of the stiffness becoming negative, and provides an experimental signature of the instability that drives domain wall motion. To confirm that this additional peak corresponds to a localized boundary mode, we perform a complementary experiment, where the domain wall remains fixed at the center site, and we measure the fluctuation response of each individual beam. From this scanning measurement, we reconstruct the mode profile using principal component analysis~\cite{han2003application, kerschen2005method}. The results reveal that the observed additional peak is indeed localized at the domain wall (Fig.~\ref{fig:experimentalImaging}\textbf{c}), and its existence can also be corroborated via finite element method simulations (Fig.~\ref{fig:experimentalImaging}\textbf{d}). 

The racetrack memory introduced here enables the construction of logic gates and circuits by forming branched networks. Figure~\ref{fig:logic}\textbf{a} illustrates a numerically simulated universal NAND gate. The gate connects the last sites of two \textquoteleft input' racetracks to the first site of an \textquoteleft output' racetrack via a negative spring. To realize NAND behavior, the output site is biased toward positive buckling, ensuring a logical 1 whenever the two inputs oppose and cancel each other. Reliable operation requires the output racetrack to be softer than the inputs; otherwise, it would perturb their dynamics. We achieve this by scaling its tight-binding parameters by 0.125. For scalability, we implement a tapered transition between input and output stiffnesses over 36 sites (Figs.~\ref{fig:logic}\textbf{b,c}). This design allows arbitrary logic circuits, as any digital circuit can be planarized using crossover gadgets~\cite{cantu2021covert}. We demonstrate this by constructing a two-bit adder (Figs.~\ref{fig:logic}\textbf{d,e}). Although these simulations are based on a tight-binding model, negative couplings and branching geometries can be physically implemented (see Ref. \cite{EmbodyingSima2025}), and the required biases are experimentally accessible (see Supplementary Information). 

Our results illustrate how a robust information transfer mechanism, based on nonlinear topological ratchets, can act as an enabler for novel forms of physical information processing~\cite{jaeger2023toward, aifer2025solving}. For example, in mechanical computers~\cite{yasuda2021mechanical, alu2025roadmap}, prior realizations of digital computations with buckling elements required re-setting the device after every iteration~\cite{mei2023memory, kwakernaak2023counting, mei2021mechanical}. In contrast, our metamaterial can operate on a continuous input stream of information. Moreover, the design introduced here is planar, and thus compatible with conventional microfabrication processes (Fig.~\ref{fig:logic}\textbf{f}), where the operation energy is extremely low~\cite{stuij2019stochastic}; in fact, at small scales, energies associated to buckling can approach the thermal scale limit $k_BT$~\cite{stuij2019stochastic}. We anticipate that these capabilities can be extended to other platforms beyond mechanics, because the response of our device is captured by a generic tight-binding model and thus describes broadly applicable physical phenomena.

\section*{Acknowledgements}

\section*{}
Correspondence can be addressed to Marc Serra-Garcia (m.serragarcia@amolf.nl) and Oded Zilberberg (oded.zilberberg@uni-konstanz.de). The authors are grateful to Martin van Hecke, Ewold Verhagen, Bernat Dura Faulí, Saeed Zohoori, Finn Bohte and Théophile Louvet for valuable inputs and discussions. 
The authors thank AMOLF’s support department: Henk-Jan Bolujit for the design and preparation of the experimental setup; Precision Manufacturing group, particularly Olaf Janssen and Mark Willemse, for fabrication of the setup and assistance with sample preparation; and Peter van Eijden for his technical assistance with the experiment.

Funded by the European Union. Views and opinions expressed are however those of the author(s) only and do not necessarily reflect those of the European Union or the European Research Council Executive Agency. Neither the European Union nor the granting authority can be held responsible for them.

This work was supported by the ERC under Grant No. 101040117 (INFOPASS). Additional funding was provided by the Deutsche Forschungsgemeinschaft (DFG) through project numbers 449653034, 521530974, and 545605411, as well as via SFB 1432 (project number 425217212). We also acknowledge support from the Swiss National Science Foundation (SNSF) through the Sinergia Grant No. CRSII5\_206008/1 and the NCCR SPIN.

\bibliography{bib}

\newpage

\appendix
\section*{Appendix}

\section{Numerical methods}

We designed the geometry of the system using the method developed in Ref. \cite{EmbodyingSima2025} through an exploration design loop. First, we extracted an effective model of masses and springs from the geometry using this method, then solved the resulting Ordinary Differential Equations (ODEs) using a 4th-order Runge-Kuta solver with 2000 points per period. To simulate variations, errors, and fabrication tolerances that naturally arise in real physical systems, we added random thermal forces to the degrees-of-freedom in the ODEs. This allowed us to evaluate the robustness of the system based on the probability of exhibiting soliton behavior in the existence of noise. We also identified the DC compression ranges (for various AC compression values) within which soliton propagation occurs. On the basis of these two evaluation criteria, we explored different values of key geometrical parameters such as the width of the sites, their distance, the type of connection, and thickness. Our final experiment corresponds to the non-dimensional values of $\lambda = 1$, $\omega_0^2 = 0.49$ , $a_{0,1} = a_{0,3} = -2.3$, $a_{0,2} = a_{0,4} = -1.0$, $a_{1,1}(\theta) = a_{1,2}(\theta) = sin(\theta)$, $a_{1,3}(\theta) = a_{1,4}(\theta) = -sin(\theta)$, and $c = 0.47$ in Eq.~\eqref{eq:GrossPitaevskii}. Although we observe that for these parameters the coupling beams can be transiently bi-stable, the propagation mechanism remains the same. To achieve these non-dimensional units, we use the scaling $q = q_0Q$, where $q_0 = \sqrt{\frac{b}{\lambda}}$ 
and $b = \max_{\theta} |a_1,j(\theta)|$, which represents the maximum amplitude of the angular-dependent coefficients. We then rescale the energy by the factor of $\frac{\lambda}{(b )^2}$.

\label{sec:figures}
\begin{figure}[tbp]
    \centering
    \includegraphics[width=\columnwidth]{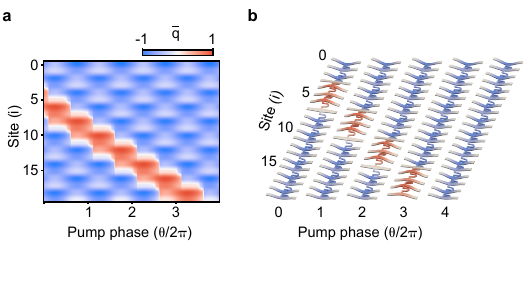}
    \caption{\textbf{Soliton propagation in the Finite Element simulation.} 
    \textbf{a} Normalized deformation fields from FEM simulation of soliton propagation, with magnitude encoded in blue-to-red color.
    \textbf{b} The 3D deformed configurations of the system at selected phases in soliton propagation as predicted by FEM.}
    \label{fig:FEM}
\end{figure}

To validate the soliton behavior of the design, we conducted FEM simulations using FEniCSx, an open-source finite element package in Python ~\cite{scroggs2022basix, alnaes2014unified, baratta2023dolfinx, scroggs2022construction}. Figure~\ref{fig:FEM} illustrates the simulated soliton propagation in different phases of the AC compression. We carried out the simulations quasi-statically and adiabatically. In each iteration, we updated the boundary conditions, solved the system statically, and used that solution as the initial condition for the next iteration. In these simulations, we employed a Saint Venant–Kirchhoff hyperelastic material model. Considering that the silicone polymer used in the experiments had a Shore A hardness of 30, we estimated the Young's modulus using the following equation ~\cite{gent1958relation}.

\begin{equation}
E = \frac{0.0981(56 + 7.62336S)}{0.137505(254 - 2.54S)}
\label{eq:Gent}
\end{equation}
where E denotes the Young's modulus in MPa, and S is the Shore A hardness defined by ASTM D2240.

For the logical gates and circuits in Fig.~\ref{fig:logic}, we use a 4th-order Runge-Kuta solver and the non-dimensional values of $\lambda = 1$, $\omega_0^2 = 1.5$ , $a_{0,1} = a_{0,3} = -2.5$, $a_{0,2} = a_{0,4} = -0.5$, $a_{1,1}(\theta) = a_{1,2}(\theta) = -sin(\theta)$, $a_{1,3}(\theta) = a_{1,4}(\theta) = sin(\theta)$, and $c = 0.5$ in the tight-binding model of Eq.~\eqref{eq:GrossPitaevskii}.

\section{Experimental methods}

\label{sec:methods}
\begin{figure}[tbp]
    \centering
    \includegraphics[width=\columnwidth]{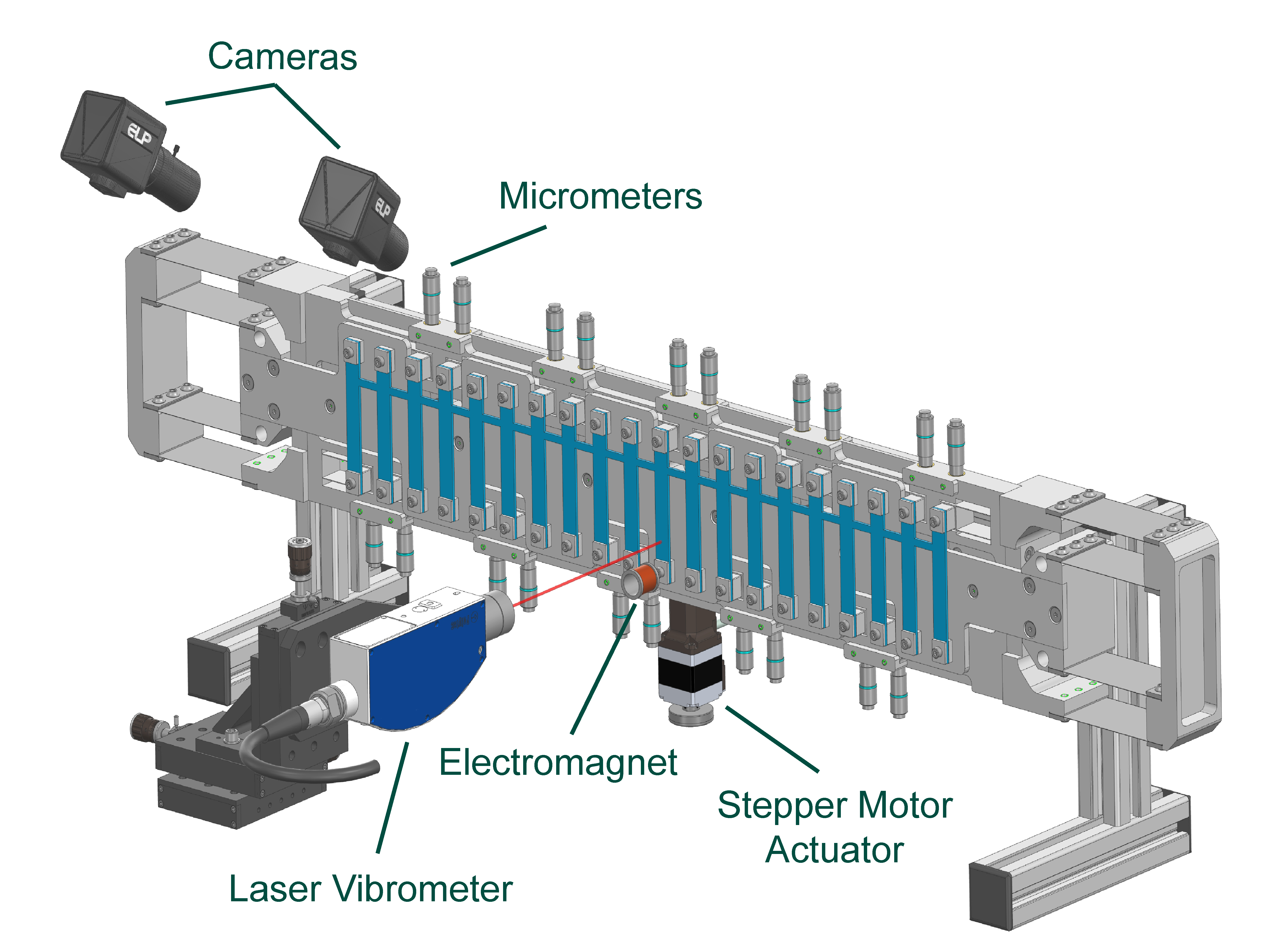}
    \caption{\textbf{Experimental setup.} Setup features a stepper motor actuator, synchronized cameras for Digital Image Correlation (DIC), micrometer heads for precise potential adjustment, an electromagnet for excitation, and a Compact Laser Vibrometer (CLV) coupled with a Digilent acquisition system.}
    \label{fig:setup}
\end{figure}

In the experimental procedure, we employed a custom-built setup consisting of different parts shown schematically in Fig.~\ref{fig:setup}. The setup comprises a Thorlabs DRV225 stepper motor actuator that enables a precise linear positioning with sub-micron resolution for the sample's AC compressive displacement, two synchronized high resolution cameras (ELP-USB4KHDR01-MFV) for capturing sample deformations, micrometer heads for fine adjustment of potential at each individual beams; an electromagnet powered by an amplifier, and excited using a Digilent Analog Discovery 2 system; and Polytec CLV-2534, a Compact Laser Vibrometer (CLV), also connected to the Digilent data acquisition system. We used the CLV to precisely measure the vibrational responses.

We prepared the samples by casting Mold Star 30 silicone rubber in a custom mold to form silicone elastomer sheets of almost 2mm thickness. This material was specifically chosen because of its low viscoelasticity and minimal plasticity. After being cured at room temperature for approximately 6 hours, we waterjet cut the resulting silicone sheets into their final geometry of 20 connected beams with a length of 105mm and a width of 10mm. Additionally, we applied speckle patterns with high-contrast coloring to the central regions of the sites to facilitate subsequent Digital Image Correlation (DIC) analysis.

\begin{figure}[!]
    \centering
    \includegraphics[width=\columnwidth]{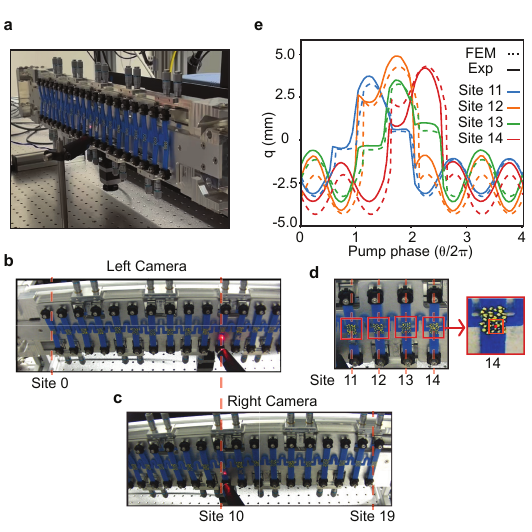}
        \caption{\textbf{Image processing overview.} 
        \textbf{a} Sample assembled on the experimental setup. 
        \textbf{b and c}, During image processing, two cameras (left and right) were used to cover all 20 sites (0 to 19). The images shown in this panel  were captured at a phase corresponding to half a cycle of the motor. 
        \textbf{d} Close-up of the corresponding sites plotted in panel d, as captured by the right camera at the -3 mm buckling configuration. The red squares approximately indicate the larger template regions used to crop the images prior to template matching, which helped reduce errors. In the inset, the red node marks the selected center point at site 14, while the orange square shows the smaller template used for template matching.
        \textbf{e} Comparison between FEM predictions and experimentally measured deformations at four selected sites, obtained through image processing. The figure demonstrates strong agreement.}
    \label{fig:DICvsExp}
\end{figure}

We assembled the samples on the experimental fixture using custom-designed supports. We aimed to minimize compressive displacement inconsistencies and promote uniform alignment and tension across all measurement sites. Precision micrometer heads allowed fine initial adjustments to site potentials, achieving a controlled initial deformation of 3 mm at the midpoint of each site, utilizing a custom-designed 3D-printed tool. Reference images were captured with all sites positioned at +3 mm (right) and -3 mm (left), providing baseline data for the image processing stage. Fig.~\ref{fig:DICvsExp}a shows the sample assembled on the setup.

Throughout the experiments, we measured deformations at the centers of the sites using a template-matching DIC routine written in Python (OpenCV 4.10). We used a 24 × 24 pixel patch cut from the reference image and matched it with the normalized cross-correlation metric of OpenCV, inside a 100 × 100 pixel search window in the current frame. The integer pixel peak was refined to subpixel precision by quadratic interpolation of the local neighborhood around the maximum. The high-contrast speckle pattern on the sample ensured accurate image matching. To capture all the beams of the sample, we used two cameras set to 1920 × 1080 pixels, as illustrated in Fig.~\ref{fig:DICvsExp}b and c. The cameras were positioned parallel to each other and faced downward as depicted in Fig.~\ref{fig:setup}.

Initially, we selected a point in the center of each site in the reference image corresponding to the -3 mm buckling configuration (Fig.~\ref{fig:DICvsExp}e). We used the image-matching algorithm to match each center to the second reference image captured at the +3 mm buckling configuration, then constructed a straight line connecting the site centers in the two positions. From this line, we obtained the undeformed center coordinates and site-specific pixel-to-millimeter scale factors. Finally, we used these parameters to convert the vertical (y-axis) image coordinates to the physical buckling values for each site at every motor step. For the soliton propagation illustrated in Fig.~\ref{fig:SystemAndPropagation}d and e we used 20 steps per period.

We used the same image correlation procedure within a feedback loop to locally adjust the boundaries and set the sites into the desired configuration, ensuring that the sample exhibited the intended soliton propagation behavior. Achieving this behavior, however, required a trial-and-error approach due to several sources of error. These sources of error included setup bias (as explained in the next section); viscoelastic and plastic effects; image correlation inaccuracies; nonuniform sample thickness and deviations from the simulated system; and assembly-related issues such as unintended residual stress and torque in the sample. Fig.~\ref{fig:DICvsExp}d compares the results of the simulated system using FEM with the deformation calculated by image processing during the experiment.

To gain a deeper understanding of the system, we captured its dynamic behavior in each phase of the motor, as illustrated in Fig.~\ref{fig:experimentalImaging}b. We conducted these measurements after each discrete and incremental movement of the motor, followed by a sufficient rest period to allow the system to reach steady state with 150 steps per period. We employed a CLV, targeting a specific region on the 10th site that was specially coated to enhance laser reflectivity. We recorded the CLV signal using a Digilent data acquisition system, sampled at 512 Hz for 8192 points, giving an FFT resolution of 0.0625 Hz. We mounted a ring magnet on the base support of the site to induce vibrational responses. The ring magnet was excited by an electromagnet connected to the Digilent system and a power source through an amplifier. The Digilent system controlled the electromagnet by providing a logarithmic chirp signal spanning 0.01–60 Hz over 16 s (512 Hz sampling, 8192 points). This type of signal helped improve the accuracy of the measurements at lower frequencies while covering a broad frequency range. Each time trace (8192 samples at 512 Hz) was transformed with an 8192-point FFT without windowing, yielding a 0.0625 Hz resolution spectrum. We retained the 0–21 Hz band and smoothed the magnitude at every motor step using a 41-point, 5th-order Savitzky–Golay filter (Savgol). The results are presented on a natural logarithmic scale. In Fig.~\ref{fig:experimentalImaging}b, we omitted one motor step in the third cycle from the fluctuation spectrum due to the insufficient reflected laser signal detected by the CLV and we replaced it with the data from the preceding step.

We conducted an additional experiment to construct the mode shape of the system, as shown in Fig.~\ref{fig:experimentalImaging}c. This experiment was carried out at a specific step of the motor where the computed frequency in ~\ref{fig:experimentalImaging}b approaches zero using the same excitation, sampling, and post-processing settings described above. We sequentially measured the vibration response of five neighboring sites on the left and right sides of the 10th site. While the excitation position remained the same, the laser moved to the corresponding site for each data point. Using the full width at half maximum (FWHM) at half power, we retained a frequency window spanning the central 70\% of the primary resonance peak of the smoothed FFT of the 10th site; the same window was applied to all other time traces. Then, we applied principal component analysis via singular-value decomposition (numpy.linalg.svd) on the resulting data matrix. The first principal component, an 11-element eigenvector, gives the spatial mode shape. 

\section{Controlling the local bias}

When a compressive force is applied to a straight beam, even small imperfections, such as asymmetric loading or geometrical defects, can significantly influence its behavior. As the axial load increases, the beam becomes increasingly susceptible to lateral deformation, since this mode of deformation requires less energy than remaining purely in axial compression, and consequently any small perturbations can cause the beam to snap into a buckled configuration. These perturbations dictate the beam's preferred instability path. 

During our experiments, we realized that the variations in the setup introduced a consistent directional bias towards positive buckling. To compensate for this, in the experiments shown in Figs.~\ref{fig:SystemAndPropagation} and ~\ref{fig:experimentalImaging} we placed the sample in an orientation in which the weaker side of the trapezoidal cross section faced the direction associated with negative buckling. This positioning served to counteract and partially mitigate the inherent bias of the setup.

\begin{figure}[tbp]
    \centering
    \includegraphics[width=\columnwidth]{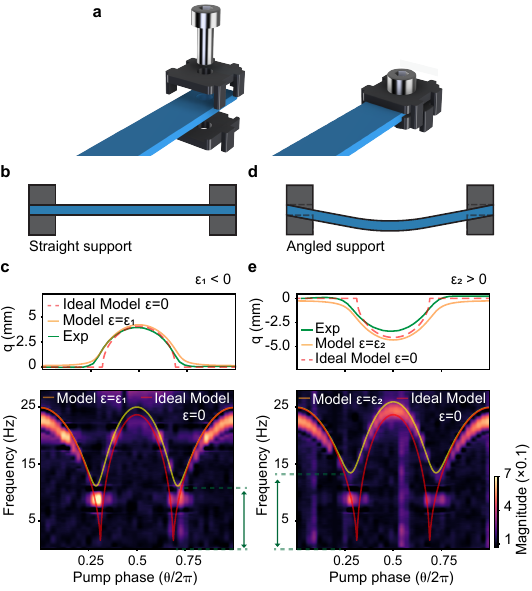}
    \caption{\textbf{Analysis of bias-induced buckling direction.}     
    \textbf{a} 3D model of the support and its assembly on the site. 
    \textbf{b} Symbolic 2D representation of a neutral support. 
    \textbf{c}, Measured Positive buckling observed in the single-sited sample using a neutral support, along with the corresponding fluctuation spectrum at different phases of the AC compressive displacement. Based on the fitted model the bias force (\(\varepsilon\)) negative. 
    \textbf{d} Symbolic 2D representation of a deliberately angled support designed to induce negative buckling.
    \textbf{e} Measured negative-induced buckling achieved using the angled support, accompanied by its corresponding fluctuation spectrum. In contrast to neutral support, here the bias force (\(\varepsilon\)) has a positive sign. One data point is omitted from the frequency plot due to the absence of a reflected signal detected by the CLV at that specific loading phase. 
    In both \textbf{d} and \textbf{e} the excitation signal from the electromagnet is a pulse. In addition, the error cause by initial axial deformation was compensated by fitting a function linearly dependent on the AC Compressive displacement.}
    \label{fig:Bias}
\end{figure}

To evaluate this directionality, we conducted a control experiment involving a single isolated site. This isolation eliminated the influence of external forces, particularly those arising from adjacent beams. As shown in Fig.~\ref{fig:Bias}b, we initially used a neutral straight support. As expected, the beam had a positive buckling configuration (see Fig.~\ref{fig:Bias}c). 

We used the following nonlinear spring equilibrium model to fit the experimental results, where \(F(x,a)\) denotes the total force acting on the beam: 

\begin{equation}
    F(q, a) = k_{1} q+k_{2} q^{3}+k_{3} q a+\varepsilon
\label{eq:fittingModel}
\end{equation}

Here, \( k_1 \), \( k_2 \), and \( k_3 \) are stiffness coefficients corresponding to the linear, Duffing-type nonlinear, and interaction terms, respectively. The variable \(q\) represents the lateral deformation of the beam at the measured point, \(a \) is the AC compressive displacement applied via the motor, and \( \varepsilon \) is a bias force representing asymmetries or imperfections in the system. 

We fitted two theoretical models to the experimental data based on equation \ref{eq:fittingModel}: one representing the idealized case without any bias (\(\varepsilon = 0\)), where the frequency approaches zero at the instability point and the beam snaps abruptly; and another incorporating a bias force into the governing equation. In the latter model, the frequency decreases with compressive displacement but does not reach zero, and the beam snaps less sharply, reflecting the influence of residual asymmetries. In this case, the bias force has a negative sign (see Fig. \ref{fig:Bias}c).

We solved the equilibrium condition \( F(q, a(\theta)) = 0 \) numerically at each step to obtain the real-valued deformation \( q_0(t) \), and computed the corresponding linearized natural frequency using the following equation.

\begin{equation}
\omega = \sqrt{\frac{1}{m} \left. \frac{dF}{dq} \right|_{q_0}} = \sqrt{\frac{k_1  + 3k_2 q_0^2 + k_3 a(\theta)}{m}}.
\label{eq:naturalFrequency}
\end{equation}

Here, \( a(t) = 1 + \cos\left( \theta \right) \) represents the periodic compressive displacement imposed by the motor. The resulting trajectories \( \omega(\theta) \) were compared directly with the experimentally measured fluctuation spectrum. The model parameters, particularly the bias term \( \varepsilon \), were manually adjusted to achieve qualitative agreement with the observed frequency softening. 

Next, we introduced a support with a deliberate angle (Fig.\ref{fig:Bias}d), which biased the system toward the negative buckling configuration (Fig.\ref{fig:Bias}e). According to the fitted model, this angled support corresponds to a positive bias force. This result highlights the system’s tunable directional bias. This feature is especially valuable for the design of mechanical logic elements, such as those illustrated in Fig.~\ref{fig:logic}a, where consistent directional bias is required. 

\section{Linearized Bogoliubov excitations band structure and nontrivial Chern number for one domain}
\label{Appendix:ChernNumber}
Here, we demonstrate that it is possible to select a pumping trajectory with a nontrivial Chern number for the fluctuation modes on top of a single infinite domain in the nonlinear metamaterial with potential $V$ [see Eq.~\eqref{eq:GrossPitaevskii}]. The corresponding Newtonian equations of motion for a particle of unit mass ($m=1$) and position $q_{n,j}$ in this potential read
\begin{align}
    0 =& \ddot{q}_{n,j} +\frac{\partial V}{\partial{q_{n,j}}}
    \\
    =& \ddot{q}_{n,j}+\lambda q_{n,j}^3+\gamma(\omega_0^2+a_{0,j}+a_{1,j}(\theta))q_{n,j}\nonumber
    \\
    & +2c q_{n,j}-cq_{l+1}-cq_{l-1}\,.\nonumber
    \label{eq:AppendixEOM}
\end{align}
We expand this equation of motion with $q_{n,j}(t) = q_{j}^{(0)} + \epsilon\delta q_{n,j}(t)$ around a stable steady state $q_{j}^{(0)}$ with 
\begin{align}
    \frac{\partial V}{\partial{q_{n,j}}}(q_{n,j}={q_{j}^{(0)}})=0\,.
\end{align}
Here, $\delta q_{n,j}$ denotes fluctuations around the steady state, and $\epsilon$ is a small parameter. We assume the steady state to consist of a single domain, such that $q_{j}^{(0)}>0$ or $q_{j}^{(0)}<0$ for all $j$. Substituting this ansatz into Eq.~\eqref{eq:AppendixEOM}, and neglecting constant energy contributions as well as terms of order $\epsilon^2$ (and higher), we obtain the fluctuation equation of motion,
\begin{align}
    0=\delta \ddot{q}_{n,j}&+(\omega_{0}^2+a_{0,j}+a_{1,j}(\theta)+2c+3\lambda (q_{j}^{(0)})^2)\delta q_{n,j}
    \\
    &- c\delta q_{l-1} -c\delta q_{l+1}\,.\nonumber
\end{align}
Equivalently, we can also formulate it in matrix form as
\begin{align}
    0 = \delta \ddot{\vec{q}} + (\mathbf{\Omega}+\mathbf{K})\delta \vec{q}
    \label{eq:AppendixFluctuationEquationOneDomain}
\end{align}
with the vector of positions $\delta \vec{q}=(...,\, \delta q_{n,j},...)$ and the effective spring constant as well as coupling matrices
\begin{align}
    \mathbf{\Omega} &= \mathrm{diag}(...,\omega_{0}^2+ a_{0,j} + a_{1,j}(\theta)+2c+3\lambda (q_{j}^{(0)})^2,...)\,\nonumber ,\\
    &=\mathrm{diag}(...,\Omega_{j}(\theta),...)\nonumber
    \\
    \mathbf{K}&=\begin{pmatrix}  
    \ddots & \ddots &  & \\  
    \ddots & 0 & -c & \\  
     &-c & 0 & \ddots\\  
    & & \ddots&\ddots   
    \end{pmatrix}\,.
\end{align}
We dub $\mathbf{D}=\mathbf{\Omega}+\mathbf{K}$ the dynamical matrix of the system. To calculate the Chern number of the linearized system, we transform to reciprocal space by writing 
$q_{n,j} = \sum_k e^{i k n} q_{k,j}$, 
where $q_{k,j}$ are the reciprocal-space coordinates and $k \in [0,2\pi]$. In this representation, the equation of motion becomes
\begin{align}
    0 &= \delta \ddot{\vec{q}}_k + \mathbf{D}_k\delta \vec{q}_k\,,\nonumber
    \\
    \mathbf{D}_k &= \begin{pmatrix}
        \Omega_1(\theta) & -c & 0 & -ce^{-\iota k}\\
        -c & \Omega_2(\theta) & -c & 0\\
        0 & -c & \Omega_3(\theta) & -c &\\
        -c e^{\iota k} & 0 & -c & \Omega_4(\theta)
    \end{pmatrix}\,,
\end{align}
with $\delta\vec q_k=(\delta q_{k,1},\delta q_{k,2},\delta q_{k,3},\delta q_{k,4})$ and the dynamical matrix $\mathbf{D}_k$ in reciprocal space.
We can now calculate the Chern number $C_r$ of the $r$-th band via
\begin{align}
    C=\frac{1}{2\pi i}\int_{0}^{2\pi}d\theta\int_0^{2\pi}dk\, (\partial_\theta A_k-\partial_k A_\theta)\, ,
\end{align}
with the Berry connection $A_\mu = \langle r(k, \theta)|\partial_\mu|r(k,\theta)\rangle$ for $\mu\in\{\theta,k\}$. We denote the normalized eigenvectors of $\mathbf{D}_k$ belonging to band $r$ with $|r(k,\theta)\rangle$. As the Chern number is only defined for two dimensional systems, we interpret the pumping parameter $\theta$ as an artificial quasimomentum.
\begin{figure}[!]
    \centering
    \includegraphics[width=\columnwidth]{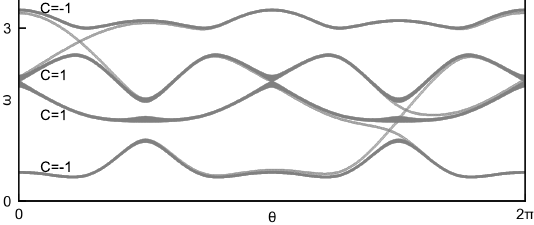}
        \caption{\textbf{Linearized band structure and Chern numbers of a single domain} 
        Linearized band structure on top of a single domain solution for $\lambda=1$, $c=0.25$, $\omega_{0}^2=2$, $a_{0,1}=-3$, $a_{0,2}=-0.5$, $a_{1,1}=-\cos(\theta)$ and $a_{1,2}=-\sin(\theta)$ realizing a clockwise circle in the $a_{1,2}$-$a_{1,1}$ plane in Fig.~\ref{fig:Fig2TheoryOfSoliton}\textbf{e}. Note, although not visible here, there is a finite band gap between the second and the third band for every $\theta$. For each band, the corresponding Chern number $C$ is annotated.}
    \label{fig:AppendixChernNumber}
\end{figure}
We can choose a pumping trajectory in the $a_{1,2}$-$a_{1,1}$ plane, such that we get nontrivial Chern numbers for the four bands, see Fig.~\ref{fig:AppendixChernNumber}. Consequently, when opening up the boundary of the system, localized edge modes that cross the band gap, appear along the pump cycle \cite{kraus2012topological,BulkEdgePumping_Hatsugai2016}. With independently controllable pumping parameters our model therefore offers the possibility of realizing a nontrivial Thouless pump of fluctuations\cite{thouless1983quantization}. Notice that in the experiment, we approximately realize a pumping trajectory where at some point $a_{1,2}$=$a_{1,1}=0$ and the band gaps close. This leads to ill-defined Chern numbers and therefore constitutes no proper Thouless pumping sequence. Nonetheless, highly localized modes at the edge of the racetrack appear throughout the pump cycle, see Figs.~\ref{fig:Fig2TheoryOfSoliton}\textbf{c} and \textbf{d}.

\section{Derivation of topological boundary mode frequency under the influence of a second domain}
\label{Appendix:ModifiedEdgeModes}
In the main text, we claim that the coupling between two domains can destabilize a localized boundary mode of one domain. In this Appendix, we show how to calculate the corresponding renormalization of such an eigenmode by applying the procedure of Ref.~\cite{theorypaper} to our classical system. 
We begin with a one-domain state. Following the steps outlined in Appendix~\ref{Appendix:ChernNumber}, we obtain the fluctuation equation [Eq.~\eqref{eq:AppendixFluctuationEquationOneDomain}] with open boundary conditions. We focus on a single domain, e.g., the red domain in Fig.~\ref{fig:Fig2TheoryOfSoliton}\textbf{a}. We simplify the notation by introducing the index $\tilde{M}=(N,1)$ to denote the first site of this domain. Diagonalizing the corresponding dynamical matrix $\mathbf{D}$ of a finite racetrack of length $M$, we obtain the eigenmodes of the system with open boundary conditions.
\begin{align}
    \mathbf{T}^{-1}\mathbf{D}\mathbf{T}=\mathrm{diag}(\omega_{1}^2, ...,\omega_{M}^2)=\tilde{\mathbf{D}}
\end{align}
with a transformation matrix $\mathbf{T}$. Even if we do not realize a pump cycle with nontrivial Chern number as in Appendix~\ref{Appendix:ChernNumber}, we can still observe strongly localized eigenmodes in the pumping pattern, see Figs.~\ref{fig:Fig2TheoryOfSoliton}\textbf{c}, \textbf{d} and Fig.~\ref{fig:experimentalImaging}\textbf{c}.

\begin{widetext}
We now turn to the case of a state with two domains, see Fig.~\ref{fig:Fig2TheoryOfSoliton}\textbf{a}. In contrast to the single-domain case, we now account for the pressure exerted by the blue domain on the red domain. As a starting point, we take the equation of motion given in Eq.~\eqref{eq:AppendixEOM}.
\begin{align}
    0 = \ddot{q}_{n,j}+\lambda q_{n,j}^3+(\omega_{0,j}^2+a_{0,j}+a_{1,j}(\theta))q_{n,j}+2c q_{n,j}-cq_{l+1}-cq_{l-1}\,.
\end{align}
Inserting the same ansatz as in Appendix~\ref{Appendix:ChernNumber}, $q_{n,j} = q_{j}^{(0)} + \epsilon \, \delta q_{n,j}$, we once again obtain a fluctuation equation. We focus on the first site of the second domain, denoted by $\tilde{M}$. Recall that in the steady state $q_{j}^{(0)}$ corresponds to a one-domain configuration. For the two-domain state, however, the solution far away from the domain wall, deep in the first domain, is given by $-q_{n,j}^{(0)}$. We assume that the first domain is unaffected by the presence of the domain wall and that its fluctuation modes remain decoupled from those of the second domain. Under these assumptions, the equation of motion for the fluctuation modes at site $\tilde{M}$ takes the form
\begin{align}
    0 &= \delta\ddot q_{\tilde{M}}+(\omega_{0,1}^2+a_{0,1}+a_{1,1}(\theta)+2c+ 3\lambda (q_{1}^{(0)})^2)\delta q_{\tilde{M}}+3\epsilon\lambda \delta q_{1}^{(0)} \delta q_{\tilde{M}}^2 +\epsilon^2\lambda \delta q_{1}^3-c\delta q_{\tilde{M}+1} + \frac{2c}{\epsilon}q_{4}^{(0)}\,.
\end{align}
In this equation, we retain all nonlinear terms. In addition, we include the coupling to the steady-state solution of the second domain, which enters proportional to $1/\epsilon$, since this contribution is not captured in the one-domain steady-state solution $q_j^{(0)}$. For site $\tilde{M}$, we keep all local nonlinearities, as these are expected to play a crucial role in determining the renormalized boundary-mode frequency. For all other sites of the domain, we linearize the equations of motion and neglect terms of order $\epsilon^2$ and higher. This yields
\begin{align}
    0 = \delta\ddot{\vec{q}}+\mathbf{D}\delta\vec q +3\epsilon\lambda  q_{1}^{(0)} \delta q_{\tilde{M}}^2 +\epsilon^2\lambda \delta q_{\tilde{M}}^3+ \frac{2c}{\epsilon}q_{4}^{(0)}\,.
\end{align}
Multiplying the diagonalization transformation for the single domain $\mathbf{T}^{-1}$ from the left leads to
\begin{align}
    0 &= \mathbf{T}^{-1}\delta\ddot{\vec{q}}+\mathbf{T}^{-1}\mathbf{D}\mathbf{T}\mathbf{T}^{-1}\delta\vec q +\mathbf{T}^{-1}(3\epsilon\lambda q_{1}^{(0)} \delta q_{\tilde{M}}^2 +\epsilon^2\lambda \delta q_{\tilde{M}}^3 + \frac{2c}{\epsilon}q_{4}^{(0)})\vec{e}_{\tilde{M}}
    \\
    0 &= \mathbf{T}^{-1}\delta\ddot{\vec{q}}+\tilde{\mathbf{D}}\mathbf{T}^{-1}\delta\vec q +\mathbf{T}^{-1}(3\epsilon\lambda q_{1}^{(0)} \delta q_{\tilde{M}}^2 +\epsilon^2\lambda \delta q_{\tilde{M}}^3 + \frac{2c}{\epsilon}q_{4}^{(0)})\vec{e}_{\tilde{M}}\,
    \\
    0 &= \ddot{\vec{w}}+\tilde{\mathbf{D}}\vec w +\mathbf{T}^{-1}(3\epsilon\lambda q_{1}^{(0)} \delta q_{\tilde{M}}^2 +\epsilon^2\lambda \delta q_{\tilde{M}}^3+ \frac{2c}{\epsilon}q_{4}^{(0)})\vec{e}_{\tilde{M}}\,.
    \label{eq:AppendixNormalModesWithLocalNonlinearity}
\end{align}
Here, we introduced the vector of eigenmodes via $\vec w=\mathbf
{T}^{-1}\delta \vec q$ and the vector $\vec{e}_{a}$, which has $M$ entries with the $a$-th being $1$ while the rest being $0$. We assume that the system has a localized boundary mode on the first site of the second domain $\tilde{M}$ with index $b$ in $\vec w$. Therefore,
\begin{align}
    (\vec{w})_b = \sum_{l=1}^{M} A_l \delta q_l\,,
\end{align}
with $|A_r|\ll |A_{\tilde{M}}|$ for every site $r\neq\tilde{M}$ in the second domain holds. Note, that $A_l=(\mathbf
T^{-1})_{b,l}$ where $b$ ($l$) indexes the row (column) of $T^{-1}$. We now perform two further approximations: (i) We  assume that the coordinate $\delta q_{\tilde{M}}$ can be identified with the boundary mode via a proportionality factor, namely
\begin{align}
    \mathbf{T}^{-1}\vec e_{\tilde{M}}\approx(\mathbf{T}^{-1})_{b,{\tilde{M}}}\vec{e}_b=A_{\tilde{M}}\vec e_b\,.
\end{align}
(ii) We assume that when representing local nonlinear terms of site $\tilde{M}$ in terms of the boundary mode, the only relevant mode-mixing is of the boundary mode itself. Therefore,
\begin{align}
\delta q_{\tilde{M}}^n\approx\Big(\frac{1}{A_{\tilde{M}}}(\vec w)_b\Big)^n\,,
\end{align}
for $n>1$. Applying these approximations to Eq.~\eqref{eq:AppendixNormalModesWithLocalNonlinearity}, we obtain
\begin{align}  
    0=\ddot{\vec{w}}+\mathbf{D}\vec{w}+\left(3\lambda q_{1}^{(0)}\left(\frac{1}{A_{\tilde{M}}}(\vec{w})_b\right)^2+\lambda\left(\frac{1}{A_{\tilde{M}}}(\vec{w})_b\right)^3+2c q_{4}^{(0)}\right)A_{\tilde{M}}\vec{e_b}\,.
\end{align}
This includes the equation of motion for the edge mode
\begin{align}
    0=\ddot{w_b} +\omega_b^2 w_b+\frac{3\lambda q_{1}^{(0)}}{A_{\tilde{M}}}w_b^2+\frac{\lambda}{A_{\tilde{M}}^2}w_b^3+2A_{\tilde{M}}\,c\, q_{4}^{(0)}\,.
\end{align}
We can linearize this equation of motion around its stable state $W_b$ to obtain the new edge mode frequency
\begin{align}
    \tilde{\omega}_b=\sqrt{\omega_b^2+\frac{6\lambda}{A_{\tilde{M}}} q_{1}^{(0)}W_b+\frac{3\lambda}{A_{\tilde{M}}^2}W_b^2}\, .
\end{align}.
We use this equation to calculate the modified boundary mode frequencies in Fig.~\ref{fig:Fig2TheoryOfSoliton}.
\end{widetext}

\end{document}